\documentclass[aps,pra,twocolumn,showpacs,floatfix,superscriptaddress]{revtex4}
\usepackage{amsmath,amssymb,amsfonts,bbm,graphicx,hyperref,times}

\def\Tr{\hbox{Tr}}
\begin{document}
\title{Optical interferometry in the presence of large phase diffusion}
\author{Marco G. Genoni}
\affiliation{QOLS, Blackett Laboratory, Imperial College London, London
SW7 2BW, UK}
\author{Stefano Olivares}
\affiliation{Dipartimento di Fisica, Universit\`a degli Studi di Milano,
I-20133 Milano, Italy}
\affiliation{CNISM, UdR Milano Statale, I-20133 Milano, Italy.}
\affiliation{Dipartimento di Fisica, Universit\`a degli Studi di
Trieste, I-34151 Trieste, Italy} 
\author{Davide Brivio}
\affiliation{Dipartimento di Fisica, Universit\`a degli Studi di Milano,
I-20133 Milano, Italy}
\author{Simone Cialdi}
\affiliation{Dipartimento di Fisica, Universit\`a degli Studi di Milano,
I-20133 Milano, Italy}
\affiliation{INFN, Sezione di Milano, I-20133 Milano, Italia}
\author{Daniele Cipriani}
\affiliation{Dipartimento di Fisica, Universit\`a degli Studi di Milano,
I-20133 Milano, Italy}
\author{Alberto Santamato}
\affiliation{Dipartimento di Fisica, Universit\`a degli Studi di Milano,
I-20133 Milano, Italy}
\author{Stefano Vezzoli}
\affiliation{Dipartimento di Fisica, Universit\`a degli Studi di Milano,
I-20133 Milano, Italy}
\author{Matteo G. A. Paris}
\email{matteo.paris@fisica.unimi.it}
\affiliation{Dipartimento di Fisica, Universit\`a degli Studi di Milano,
I-20133 Milano, Italy}
\affiliation{CNISM, UdR Milano Statale, I-20133 Milano, Italy.}
\date{\today}
\date{\today}
\begin{abstract}
Phase diffusion represents a crucial obstacle towards the implementation
of high precision interferometric measurements and phase shift based
communication channels.  Here we present a nearly optimal
interferometric scheme based on homodyne detection and coherent signals
for the detection of a phase shift in the presence of large phase
diffusion. In our scheme the ultimate bound to interferometric
sensitivity is achieved already for a small number of measurements, of
the order of hundreds, without using nonclassical light.
\end{abstract}
\maketitle
\section{Introduction}
Optical interferometry represents a high accurate measurement scheme
with wide applications in many fields of science and technology
\cite{Cav81,r1,r2,r3,r4}.  
Besides, the precise estimation of an optical phase shift is relevant 
for optical communication schemes where information is encoded in the 
phase of travelling pulses. Several experimental protocols have been
proposed and demonstrated to estimate the value of the optical phase  
\cite{Armen02,Mitch04,Nag07,Res07,Hig07,Hig09} and
showing the possibility to attain the so-called Heisenberg limit
\cite{Z92,Sam92,Lane93,San95,Eck06,Giov046,Guo08,Sme08,Gro11,Hay11,Bra11}. 
Recent developments also revealed the potential 
advantages of nonlinear interactions \cite{Boi08}.
However, in realistic conditions, 
one has to retrieve phase information that has been unavoidably degraded 
by different sources of noise, which have to be taken into account in order 
to evaluate the interferometric precision \cite{Gio11}.
The effects of imperfect photodetection in the measurement stage, or the
presence of amplitude noise in the interferometric arms have been
extensively studied 
\cite{Par95,Cam03,Huv08,Coo10,BanPhNat,BanPhPRL,Cab10,Joo11,DurkinPh,Bah11,Dat11}. 
Only recently, the role of phase-diffusive noise in interferometry have
been theoretically investigated for optical polarization qubit
\cite{Bri10,Ber10,tes:11}, condensate systems \cite{BEC1,BEC2}, Bose-Josephson
junctions \cite{Fer10}, and Gaussian
states of light \cite{Gen11}.  As a matter of fact, phase-diffusive
noise is the most detrimental for interferometry  and any signal that is
unaffected by phase-diffusion, is also invariant under a phase shift,
and thus totally useless for phase estimation. 
\par
In this paper, we present an experimental interferometric scheme
where phase diffusion may be inserted in a controlled way, and
demonstrate that homodyne detection and coherent signals are nearly 
optimal for the detection of a
phase shift in the presence of large phase diffusion. 
Indeed, while in ideal conditions squeezed vacuum is the 
most sensitive Gaussian probe state for a given average photon
number \cite{Mon06}, for large phase-diffusive noise, 
coherent states become the optimal choice, outperforming
squeezed states \cite{Gen11}.
In our scheme the ultimate bound to interferometric sensitivity, as dictated 
by the Cram\'er-Rao (CR) theorem, is achieved already for a small number of
repeated measurements, of the order of hundreds, using Bayesian
inference on homodyne data and without the need of nonclassical light.
\par
The paper is structured as follows: In Section \ref{s:intn} we describe
the evolution of a light beam in a phase diffusing environment as well
as the bound to interferometric precision in the presence of phase
noise. In Section \ref{s:expa} we describe our experimental apparatus,
whereas the experimental results are reported and discussed in Section 
\ref{s:expr}. Section \ref{s:outro} closes the paper with some concluding remarks.
\section{Interferometry in the presence of phase diffusion}
\label{s:intn}
The evolution of a light beam in a phase diffusing environment 
is described by the master equation $$\dot{\varrho} = \Gamma 
\mathcal{L}[a^{\dag} a] \varrho\,,$$ where $\mathcal{L}[O]\varrho 
= 2 O \varrho O^\dag - O^\dag O \varrho - \varrho O^\dag O$
and $\Gamma$ is the phase damping rate.  An initial state 
$\varrho_0$ evolves as $$ \varrho_t = \mathcal{N}_{\Delta} (\varrho_0)  
= \sum_{n,m} e^{- \Delta^2 (n-m)^2} \varrho_{n,m} |n\rangle\langle m|\,,$$
where $\Delta\equiv \Gamma t$, and $\varrho_{n,m}= \langle n | \varrho_0|m\rangle$.
The diagonal elements are left unchanged, in fact
energy is conserved, whereas the off-diagonal ones are progressively 
destroyed, together with the phase information carried by the state. 
Phase diffusion corresponds to the application of a random, zero-mean
Gaussian-distributed phase shift, i.e.,
\begin{align}
\varrho_{t} = \int_{\mathbbm R}\! d\beta \,
g(\beta|\Delta)
U_\beta \varrho_0 U_\beta^{\dag} \label{g:noise}
\qquad g(\beta|\Delta)=\frac{e^{-\beta^2/(4 \Delta^2)}}{\sqrt{4 \pi \Delta^2}}
\end{align}
where 
$U_\beta=\exp\{-i \beta (a^\dag a) \}$ is the phase shift
operator.
\par
We assume that the phase noise occurs between the application
of the phase shift and the detection of the signal, and consider
the estimation of a phase shift applied to a single-mode
coherent state. Homodyne detection is then performed on the output state
$$
\varrho_{\Delta, \alpha}(\phi) = \mathcal{N}_{\Delta} (U_\phi
|\alpha\rangle \langle\alpha | U_\phi^\dag) \,,$$ and the value of the
unknown phase shift $\phi$ is inferred using Bayesian estimation
applied to homodyne data.  Notice, however, that since the phase noise
map and the phase shift operation commute, our results are valid also
when the phase shift is applied to an already phase-diffused coherent
state.  The precision of the above procedure is then compared with the
benchmarks given by i) the quantum CR bound for coherent states and
any quantum limited kind of measurement, ii) the ultimate precision
achievable with optimized Gaussian states, i.e., the quantum CR bound
for general Gaussian signals, where, e.g., we allow for squeezing.
\subsection{Interferometric precision in the presence of phase
noise}
The quantum CR bound \cite{Mal9X,BC9X,Bro9X,LQE,Dav11} is obtained
starting from the Born rule $p(x|\phi)= \Tr[\Pi_x \varrho_\phi]$ where
$\{\Pi_x\}$ is the operator-valued measure describing the measurement
and $\varrho_\phi$ the density operator of the family of phase-shifted
states under investigation.  Upon introducing the (symmetric)
logarithmic derivative $L_\phi$ as the operator satisfying
$$2 \partial_\phi\varrho_\phi= L_\phi \varrho_\phi+ \varrho_\phi
L_\phi\,,$$ one proves that the ultimate limit to precision (independently
on the measurement used) is given by the quantum CR bound 
$$\textrm{Var}(\phi) \geq [M H(\phi)]^{-1}\,,$$ where $H(\phi)=
\hbox{Tr}[\varrho_\phi\,L_\phi^2]$ is the quantum Fisher information (QFI). 
The ultimate sensitivity of an
interferometer thus depends on the family of signals used to probe the
phase shift and thus, as said above, we are going to compare the
precision of our interferometer with the maximum achievable with
coherent states, and with the ultimate precision achievable with
optimized Gaussian states (for more details about the derivation of
the corresponding quantum CR bounds see \cite{Gen11}).
\par
Homodyne detection measures the field quadrature $$x_\theta=\frac12 (a
e^{-i\theta}+a^\dag e^{i\theta})\,,$$ where $\theta=\arg\alpha + \pi/2$ is
set to the optimal value to detect the imposed phase shift.  The
likelihood of a set of homodyne data $$X=\{x_1,x_2,\ldots,x_M\}\,,$$ is
the overall probability of the sample given the unknown phase $\phi$,
i.e., $$ L(X|\phi) = \prod_{k=1}^{M} p(x_k|\phi)\,,$$
where 
\begin{align}
p (x |\phi) = \frac{ e^{-2 x^2}}{\pi \Delta} 
\int_\mathbbm{R}\!\!d\beta \: 
e^{-\frac{\beta^2}{2 \Delta^2}
+ 4\alpha x \cos(\beta+\phi) -2 \alpha^2 \cos^2(\beta + \phi)}\,.
\notag
\end{align}
Assuming that no a priori information is available on  
the value of the phase shift (i.e., uniform prior), 
and using the Bayes theorem, 
one can write the {\em a posteriori} probability 
\begin{align}
P(\phi|X) = \frac{1}{\cal N}\, L(X|\phi) \qquad {\cal N} 
= \int_{\Phi} d\phi\,L(\phi|X)\,,
\end{align}
$\Phi=[0,\pi]$ being the parameter space. The probability $P(\phi|X)$
is the expected distribution of $\phi$ {\em given} the data sample
$X$. The Bayesian estimator $\phi_{\rm B}$ is the mean of the a
posteriori distribution, whereas the sensitivity of the overall
procedure corresponds to its variance
$${\rm Var}[\phi_{\rm B}] =
\int_{\Phi}d\phi\,(\phi-\phi_{\rm B})^2\,P(\phi|X)\,.$$
Bayesian estimators are known to be asymptotically unbiased and
optimal, namely, they allow one to achieve the CR bound as the size of
the data sample increases \cite{H9X,O09}. On the other hand, the
number of data needed to achieve the asymptotic region may depend on
the specific implementation \cite{Bar00}. In the following we will
experimentally show that our setup achieves optimal estimation already
after collecting few hundreds of measurements.
\section{Experimental apparatus}
\label{s:expa}
\begin{figure}[h!]
\includegraphics[width=0.99\columnwidth]{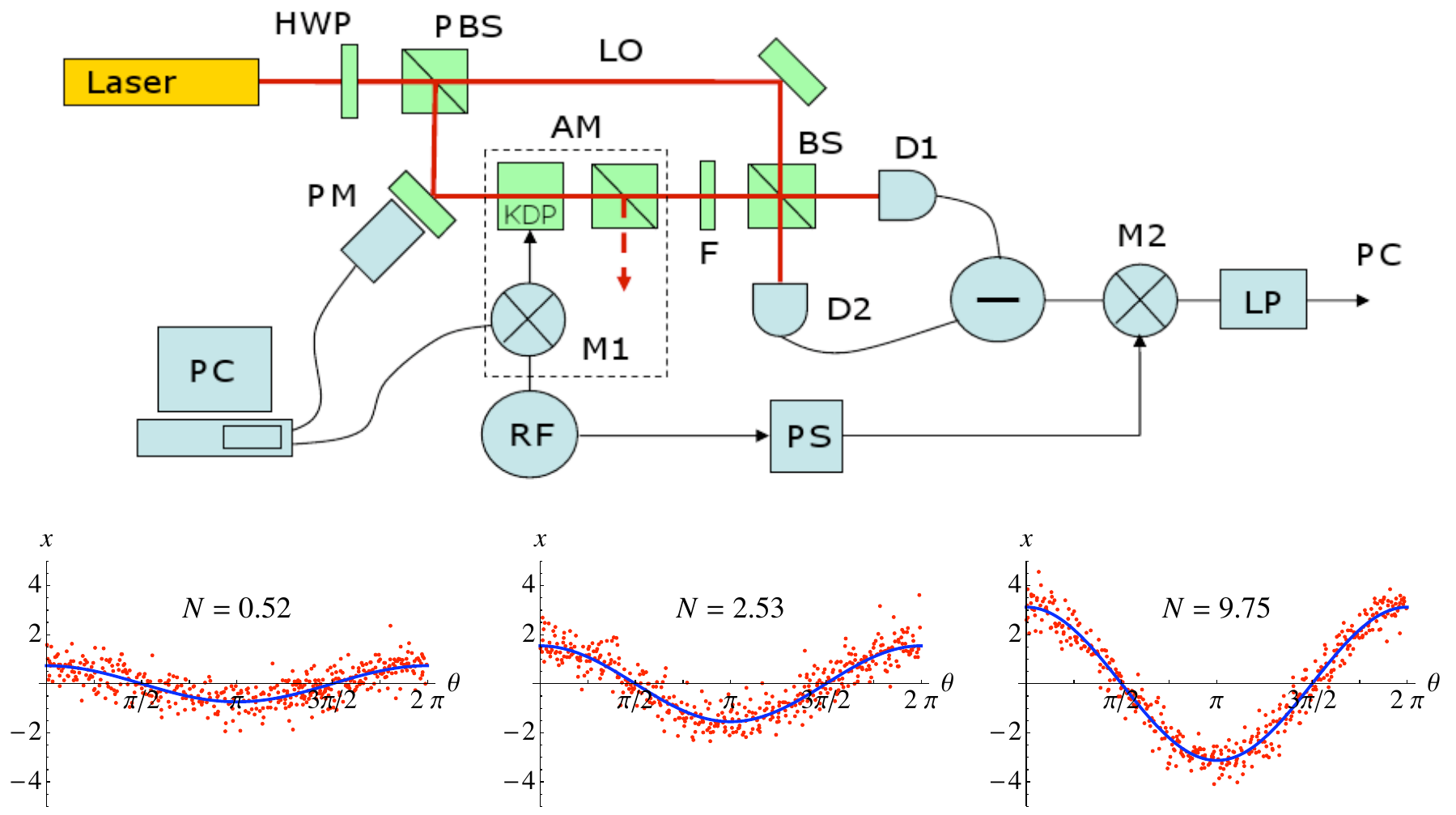}
\caption{(color online). Schematic diagram of the experimental setup. A He:Ne
laser is divided into two beams, one acts as the local
oscillator and the other represents the signal beam. The signal is
modulated at $4$MHz  with a defined modulation depth to control the
average number of photons in the generate state. One of the mirrors in the signal
path is piezo mounted to obtain a variable phase difference between the
two beams. The data are recorded by a homodyne detector
whose difference photocorrent is demodulated and then acquired by a
computer after a low pass filter. We also show the 
typical homodyne samples obtained for coherent signals of 
different amplitudes by varying the phase of the local oscillator 
(these are used to check the calibration of the piezo, which is
performed using signals with a larger number of photons).}
\label{exp_setup} \end{figure}

A schematic diagram of the interferometer is reported in
Fig.~\ref{exp_setup}. The principal radiation source is provided by a
He:Ne laser (12~mW, 633~nm) shot-noise limited above 2~MHz. The laser
emits a linearly polarized beam in a TEM00 mode. The beam is splitted
into two parts of variable relative intensity by a combination of a
halfwave plate (HWP) and a polarizing beam splitter (PBS). The
strongest part is sent directly to the homodyne detector where it acts
as the local oscillator, whereas the ramaining part is used to encode
the signal and will undergo the homodyne detection. The optical paths
travelled by the local oscillator and the signal beams are carefully
adjusted to obtain a visibility typically above 90\% measured at one
of the homodyne output ports. The signal is amplitude modulated at
4~MHz with a defined modulation depth to control the average number of
photons in the generate state. 
\par
The amplitude modulation system consist
of a KDP non-linear crystal with the $xy$ axes at 45$^\circ$, and a
PBS. The modulation is applied at the KDP crystal by means a waveform
generator Rohde \& Schwarz and a power amplifier Mini-Circuits
ZHL-32A.  The modulation depth is imposed at the proper level by a
computer that sends a costant voltage to a mixer (M1) located between
the waveform generator and the power amplifier.  One of the mirrors in
the signal path is piezo mounted to obtain a variable phase difference
between the two beams. The piezo is preloaded and its resonance
frequency is 13.5~kHz.  
\par
The phase difference is controlled by the
computer after a calibration stage. The computer sends a voltage
signal between 0 and 10~V that corresponds at the phase diffusion with
a frequency of 5~kHz to a power amplifier based on LM675 integrated
circuit that is able to drive the piezo at this frequency. With this
system it is possibile to generate any kind of phase modulation.
 \par
 The detector is composed by a 50:50 beams splitter (BS) and a
 balanced amplifier detector with a bandwidth of 50~MHz. The
 difference photocurrent is filtered with high pass filters, amplified
 and demodulated at 4~MHz by means of an electrical mixer (M2).  In
 this way the detection occurs outside any technical noise and, more
 importantly, in a spectral region where the laser does not carry
 excess noise.  The signal is filterd by a low pass filter with a
 bandwidth of 300~kHz and sent to the computer through the National
 Instrument multichannel data acquisition 6251 with 16~bit of
 resolution and 1.25~MS/s sampling rate.  The same device is used to
 send diffusion parameters to the phase modulator and signal
 parameters to the amplitude modulator.
\section{Experimental results}
\label{s:expr}
In this Section, we present our experimental results, obtained 
with signals of different energies and different levels of noise.
At first we show homodyne samples with the corresponding {\em a
posteriori} distributions and then compare the precision obtained
in our scheme with the ultimate bound imposed by the (quantum) 
Cram\'er-Rao theorem. Finally, we analyze the dependence of precision
on the signal energy and the noise in order to illustrate how in the 
limit of large phase diffusion coherent states becomes 
the optimal Gaussian probe states. In fact, they outperform
squeezed vacuum states, whose non-classical features are degraded by 
phase diffusion process, to an extent that make them useless 
for quantum metrology.
\par
\begin{figure}[h!]
\includegraphics[width=0.99\columnwidth]{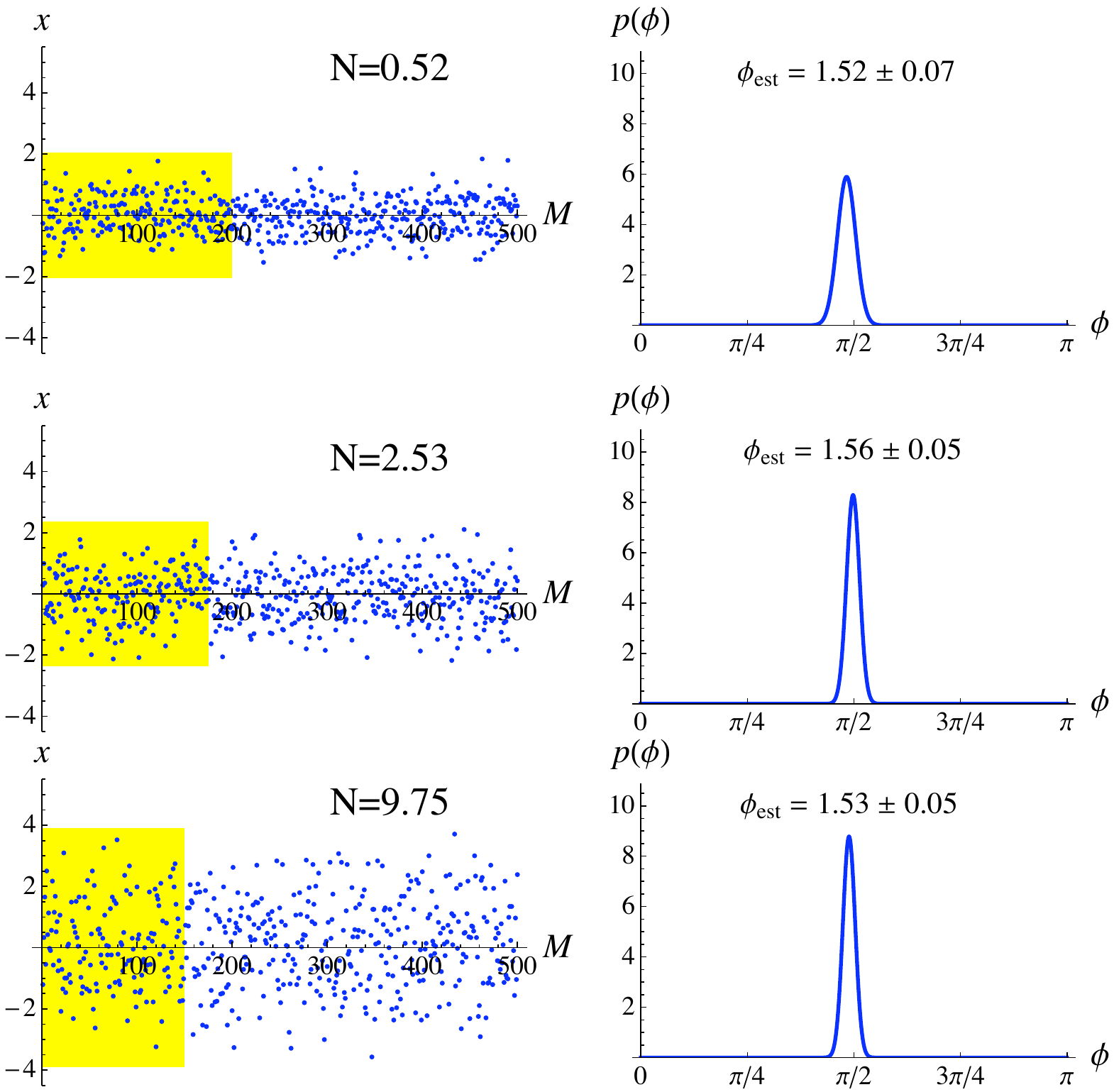}
\caption{(color online)
\label{f:inf} 
Typical examples of homodyne samples measured at fixed optimal $\theta$, 
together with the corresponding  Bayesian {\em a posteriori} distribution 
for the phase shift. The phase diffusion is $\Delta=\pi/6$ rad and the 
yellow area denotes the portion of data used to infer the phase shift.
}
\end{figure}

In Fig.~\ref{f:inf} we report typical examples of homodyne samples,
referred to a coherent signal with $N=|\alpha|^2$ mean photon number
measured at fixed optimal $\theta$, together with the corresponding
Bayesian {\em a posteriori} distribution for the phase shift. The
yellow area denotes the portion of data used to infer the phase shift.
We choose this range in order to emphasize that the optimality
region in achieved already in that region.
In fact, upon considering larger samples, precision would be
improved, due to the statistical scaling of the variance ${\rm
  Var}[\phi] = C/M$, $C$ being a proportionality constant. On the
other hand, optimality, i.e., the fact that 
$$
C\simeq 1/H_{\alpha}\,,
$$ 
{where $H_{\alpha}$ is the QFI for phase-diffused coherent signals, 
is  achieved for $M\sim 100$ measurements. In the noiseless 
case the QFI is given by $H_\alpha =4N$, whereas it decreases monotonically
by increasing the value of the noise parameter $\Delta$. 
Notice that using  optimized Gaussian signals, i.e. the 
squeezed vacuum state, one has a QFI given by $H_g = 8 N^2 + 8 N$ in the noiseless 
case. However, in the presence of large phase diffusion, i.e. for 
large values of $\Delta$, $H_\alpha$ is larger than the QFI 
obtained for phase-diffused squeezed 
vacuum states. In other words, coherent states turns out to be the 
optimal Gaussian probe states \cite{Gen11}}.
\par
\begin{figure}[h!]
\includegraphics[width=0.99\columnwidth]{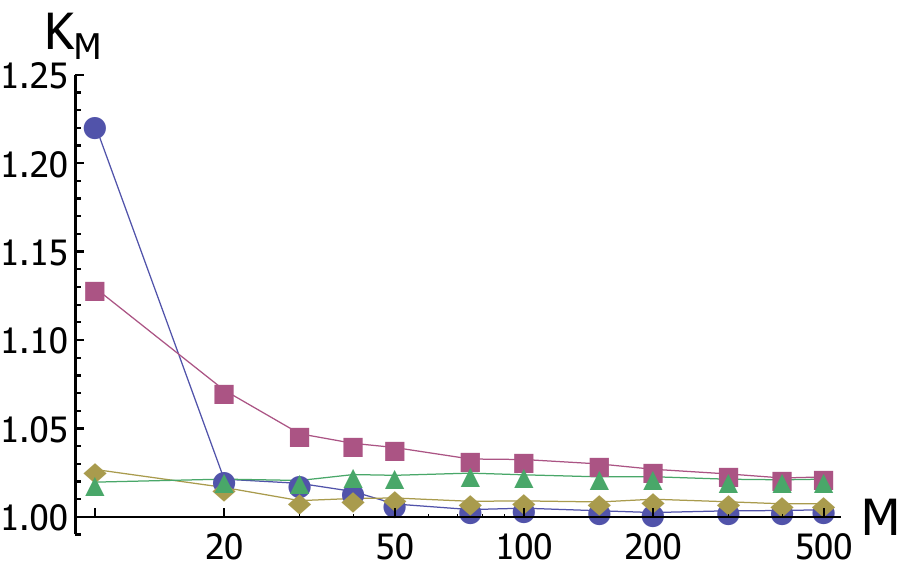}
\caption{(color online)
\label{f:varyingM} The noise ratio $K_M=({\rm Var}[\phi_{\rm B}] M H_\alpha)$ 
as a function of the number of data $M$ and for different values of 
the number of photons $N$ and the noise parameter $\Delta$. 
Blue circles: $N=0.90$, $\Delta=\pi/18$ rad;
red squares: $N=0.90$, $\Delta=\pi/9$ rad; 
yellow diamonds: $N=4.12$, $\Delta=\pi/18$ rad;
green triangles: $N=4.12$, $\Delta=\pi/9$ rad.}
\end{figure}

In Fig.~\ref{f:varyingM} we plot the quantity $$K_M = M\, {\rm
  Var}[\phi_{\rm B}] H_\alpha\,,$$ i.e., the variance of the Bayesian
estimator from homodyne data multiplied by the number of data
(measurements) and by the coherent states quantum Fisher information,
as a function of $M$. $K_M$ is by definition larger than one and
expresses the ratio between the actual precision of the
interferometric setup and the CR bound.  As it is apparent from the
plot $K_M$ rapidly decreases with the number of measurements, almost
independently on the value of the number of photons $N$ and of the
noise parameter $\Delta$. The optimality region, i.e., $K_M\simeq 1$
is achieved already for $M\simeq 100$ measurements, and the asymptotic
value of $K_M$ is closer to 1 for increasing $N$ and
$\Delta$. Furthermore, the number of measurements needed to achieve
the optimal region may be (slightly) reduced by using the Jeffreys
prior \cite{Jpp} $$p(\phi) \propto \sqrt{F(\phi)}\,$$ instead 
of the uniform one, where
$$F(\phi)=\int\!dx\, p(x|\phi)[\partial_\phi \log p(x|\phi)]^2$$ is the
Fisher information of the homodyne distribution. \par 
\par
\begin{figure}[h!]
\includegraphics[width=0.99\columnwidth]{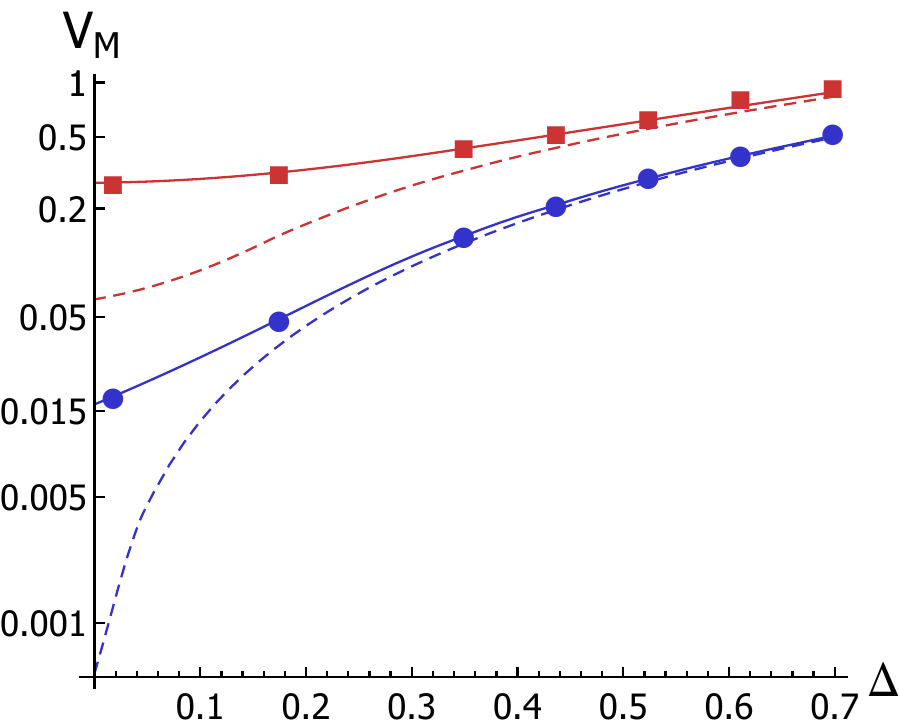}
\includegraphics[width=0.99\columnwidth]{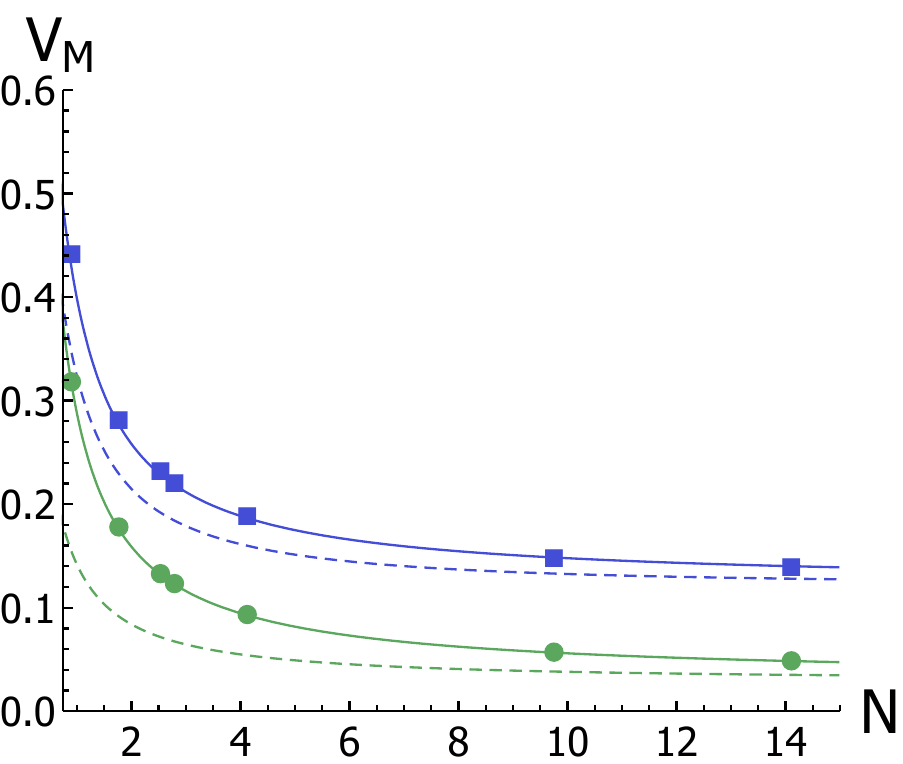}
\caption{(color online)
\label{f:varDN} Variance $V_M=M{\rm Var}[\phi_{\rm B}]$ 
of the Bayesian estimator from homodyne data after $M=100$
measurements (points), together with the CR bound
for coherent states (solid lines) and for optimized Gaussian 
states (dashed lines). The top panel of shows the behaviour 
of $V_{100}$ as a function of $\Delta$ 
for different values of the number of photons 
(top red lines/squares: $N=0.90$;  bottom blue lines/circles: $N=14.11$).
The bottom panel shows $V_{100}$ as a function of the 
number of photons $N$ and for different
values of the noise (top blue lines/squares: $\Delta=\pi/9$ rad;  
bottom green lines/circles: $\Delta=\pi/18$ rad).}
\end{figure}

In Fig.~\ref{f:varDN} we show the variance of the Bayesian estimator
from homodyne data $$V_M=M{\rm Var}[\phi_{\rm B}]$$ obtained after $M$
measurements, together with the CR bound $1/H_\alpha$ for coherent
states, and for the (phase-diffused) optimized Gaussian states, i.e., 
$1/H_g$. In particular, the top panel shows the behaviour as a function of
$\Delta$ for different values of the number of photons $N$, while in
the bottom panel we plot the same quantities as a function of the
number of photons $N$ and for different values of the noise
$\Delta$. As it is apparent from the plots, nearly optimal
inferferometric precision is achieved for increasing energy or phase
diffusion, i.e., for larger values of $N$ or $\Delta$.
\section{Conclusions}
\label{s:outro}
In conclusion, we have demonstrated a nearly optimal 
interferometric scheme based on homodyne detection and 
coherent signals for the detection of a phase shift in 
the presence of large phase diffusion. Our scheme does not 
require nonclassical light and achieve the ultimate bound to 
interferometric sensitivity using Bayesian analysis on 
small samples of homodyne data, where the number of measurements 
is of the order of few hundreds. \par
It is worth noting that for large phase diffusion coherent states 
are the optimal Gaussian probe states. Indeed they outperform 
squeezed vacuum states, whose non-classical features are degraded by 
phase diffusion process, such that they become completely useless 
for quantum metrology.
\par
Optical interferometry represents a high accurate measurement scheme
with wide applications in many fields of science and technology,
including high precision measurements and communication channels.  On
the other hand, phase diffusion represents a crucial obstacle towards
the implementation of high precision interferometric measurements and
phase shift based communication channels.  Our results allow to design
feasible, high-performance, communication channels also in the presence
of phase noise, which cannot be effectively controlled in realistic
conditions.  Therefore, besides fundamental interest, our results also
represent a benchmark for realistic phase based communication or
measurement protocols.
\par
\par
\section*{Acknowledgements}
This work has been supported by MIUR (FIRB ``LiCHIS'' - RBFR10YQ3H), the 
UK EPSRC (EP/I026436/1), MAE (INQUEST), UniMi (PUR2009 SIN.PHO.NANO), 
UIF/UFI (Vinci Program), and the University of Trieste (FRA2009).

\end{document}